\documentclass[aps,prd,preprint,superscriptaddress,showpacs,nofootinbib]{revtex4-1}

\usepackage{graphics}
\usepackage{epsfig}
\usepackage{latexsym}
\usepackage{colordvi}
\usepackage{amsmath}
\usepackage{amssymb}
\usepackage{slashed}

\usepackage[title]{appendix}

\begin{document}

\preprint{\parbox[b]{1in}{ \hbox{\tt PNUTP-21/A01}  }}
\preprint{\parbox[b]{1in}{ \hbox{\tt APCTP Pre2021-23}  }}

\title{The effects of magnetic fields on magnetic dipole moments}

\author{Gyurin Kim}
\affiliation{Department of Physics, Pusan National University,
             Busan 46241, Korea}

\author{Tuna Demircik}
\email{tuna.demircik@apctp.org}
          
\affiliation{Asia Pacific Center for Theoretical Physics, Pohang, 37673, Korea}
\author{Deog Ki Hong}
\email[E-mail: ]{dkhong@pusan.ac.kr}

\affiliation{Department of Physics, Pusan National University,
             Busan 46241, Korea}
             
\author{Matti J\"arvinen}
\email{matti.jarvinen@apctp.org}            
\affiliation{Asia Pacific Center for Theoretical Physics, Pohang, 37673, Korea}
\affiliation{Department of Physics, Pohang University of Science and Technology, Pohang, 37673, Korea}

\vspace{0.1in}

\date{\today}

\begin{abstract}
We calculate the effect of magnetic fields on the magnetic dipole moment of leptons up to the quadratic order in the magnetic field, including the QCD contributions. Since the leading contribution, which is linear in the magnetic field, depends on the spin, its effect is not measurable directly in the Penning trap experiment for the electron dipole moment. In the muon anomaly, however, we find that the electrons decayed from muons are refracted linearly in the magnetic field. This effect, though quite small, changes the distribution of the detected electrons in the muon $g-2$ experiment and could be measurable, improving the experimental uncertainties. 
  We also discuss the general field-dependent form factors and the Ward-Takahashi identity under the external magnetic fields.\end{abstract}


\maketitle

\newpage

\section{Introduction}

Magnetic moments of leptons have provided one of the most stringent tests for quantum field theory (QFT) since its
inception~\cite{Schwinger:1951nm}. They were 
later conveniently used to test the gauge independence of the S-matrix in the standard electroweak theory during 
its 
establishment~\cite{Fujikawa:1972fe}. Currently they offer 
an intriguing possibility for new physics.
QFT predicts that
the gyromagnetic ratio of the magnetic moments  
deviates from the value obtained by Dirac, $g_l=2$, due to the radiative corrections such as the self energy and the vacuum polarizations. In the case of electrons the estimate of the anomalous magnetic moment, $a_e\equiv(g_e-2)/2$ 
(i.e., 
the deviation from the Dirac value in 
units  
of Bohr magneton, $\mu_B$) 
obtained by using  
the standard model (SM) of particle physics, agrees with the experimental measurement based on a Penning trap at the level of the parts per billion (ppb)~\cite{Hanneke:2010au}. Similarly the anomalous magnetic moment of muon has been accurately measured at the level of parts per million (ppm) at Brookhaven National Laboratory (BNL)~\cite{Muong-2:2006rrc} in an experiment utilizing the difference in the cyclotron frequency and the spin precession frequency of muon under an external magnetic field. 

Recently the Fermilab Muon $g-2$ Experiment, which is an improved version of the previous BNL experiment, published their first results, 
consistent with the BNL Muon E821 measurement~\cite{Muong-2:2021ojo}. Combining these two experiments, one finds a $4.2\,\sigma$  discrepancy between the SM estimate and the experimental measurement, indicating a possible new physics effect on the magnetic moment of muon. 
While it is intriguing to attribute the discrepancy in the muon magnetic anomaly, $a_{\mu}\equiv(g_{\mu}-2)/2$, to the physics beyond SM, certain caution needs to be exercised until we have a good control of the known-physics contributions to the anomalous magnetic moment. Interestingly the recent lattice calculation by the BMW collaboration on the leading hadronic contribution to the muon magnetic moment, which is known to have the largest theoretical error~\cite{Davier:2019can}, 
suggests 
that the previous estimate, based on the dispersion relation,  somewhat underestimates the leading hadronic contribution~\cite{Borsanyi:2020mff}. If one assumes the recent BMW lattice 
result, whose estimate on the $R$-ratio is though about $2\,\sigma$ larger than the experimental value, the discrepancy in the muon anomaly reduces to $3\,\sigma$, much smaller than the threshold for an experimental discovery.

In this paper, among possible known-physics, we focus on the effect of magnetic fields and discuss how it could change the 
results obtained 
in the experiments that measure the magnetic moment of electrons and muons.  As the experimental accuracy approaches the level of ppb for the electron anomaly, the effect of the finite magnetic fields may not be negligible.   One expects by the dimensional analysis 
the correction due to magnetic fields on the magnetic moment anomaly to be roughly given as
\begin{equation}
\delta (g-2)\sim \frac{\alpha}{2\pi}\cdot\frac{eB}{m^2}=3\times 10^{-13}\left(\frac{B}{10\,{\rm kG}}\right)\left(\frac{0.51\,{\rm MeV}}{m}\right)^2\,,
\end{equation}
which is comparable to the current experimental uncertainty, $\delta a_e^{\rm exp}\simeq 2.8\times 10^{-13}$, for electrons~\cite{Hanneke:2008tm}, though it is much smaller for the experimental uncertainty in the muon anomaly, $\delta a_{\mu}^{\rm exp}\sim 10^{-10}$ of the current Fermi Lab experiment.

The magnetic dipole moment is an intrinsic quantity of elementary particles. Its operational definition is how they react to the external magnetic fields in the limit where the magnetic fields vanish: For example the magnetic moment of a particle of mass $m$ and  electric charge $q$ with spin $\vec S$ could be measured from the change of the energy by the external magnetic fields, taken to be vanishingly small, as
\begin{equation}
\vec \mu\equiv g\frac{q\vec S}{2m}=-\left.\frac{\partial {\cal E}(\vec B)}{\partial \vec B}\right|_{\vec B=0}\,.
\end{equation}
But in practice  the magnetic fields never vanish. In fact one often needs a strong magnetic field, about $10\,{\rm kG}$ or higher  for the measurement. Hence one may not neglect the effect of magnetic fields to achieve an extreme accuracy in the measurements of magnetic dipole anomalies of electrons. In the case of muons,   the direct effect of magnetic fields on them is quite small though, since they are much heavier than electrons. However, as the electrons created under the magnetic field from the decay of muons are measured to obtain the muon anomaly in the current experiments,  the magnetic field effect could be in principle comparable to that of electrons. 

The remaining of the paper is organized as following: In Section~\ref{sec2} we present our calculations on the magnetic field-dependent dipole moment up to the quadratic order in the field, including both QED and QCD contributions. In section~\ref{sec3} we discuss how the electrons are refracted due to the magnetic field and then we analyze the refraction effect on the measurement of the muon magnetic anomaly.  In section~\ref{con} we make our conclusions. Finally in the appendix we present the general (magnetic) field-dependent CP-even form factors and we prove the Ward-Takahashi identity in the presence of magnetic field, needed to check the consistency of our calculations.   
\section{Field-dependent magnetic moment}
\label{sec2}
To measure the magnetic moment of electron,  one normally confines a cloud of free electrons in a cylindrical Penning trap using a magnetic field along the trap axis $\vec B=B\,\hat k$ with an additional electrostatic quadrupole potential~\cite{penning,penning2}. By measuring the energy difference of the so-called perturbed cyclotron mode of electrons in the trap, defined as $\left|\Delta {\cal E}\right|\equiv g_e\mu_B B$, due to the spin flip, one gets the precise value of the gyromagnetic ratio $g_e$ of electrons. 
Since, however,  by quantum fluctuations the electrons will be scattered many times by the external magnetic field, the energy difference will be in general a function of the external magnetic field, $B$, leading to the field-dependent $g$ factor, which can be expanded for the weak magnetic fields as
\begin{equation} \label{ge_def}
g_e(B)\equiv\frac{\left|\Delta {\cal E}\right|}{\mu_BB}=g_e(0)+\sum_{n=1}^{\infty}g_e^{(n)}\left(\frac{eB}{m^2}\right)^n\,.
\end{equation}  

\subsection{Field dependence in QED}

In QED the exact electron propagator to the order $\alpha$ under a constant magnetic field has been derived by Schwinger~\cite{Schwinger:1951nm}, from which the shift in the ground state energy due to the radiative corrections is found to be at one-loop~\cite{Newton:1954zz,Tsai:1973jex}
\begin{equation}
\Delta{\cal E}^{\rm QED}(B)=\frac{\alpha}{2\pi}m\left[-\frac{eB}{2m^2}+a_2\left(\frac{eB}{m^2}\right)^2+a_3\left(\frac{eB}{m^2}\right)^3+\cdots\right]\,,
\label{qed_e}
\end{equation}
where $a_2=\left(\frac43\ln\frac{m^2}{2eB}-\frac{13}{18}\right)$, 
and
$a_3=\left(\frac{14}{3}\ln\frac{m^2}{2eB}-\frac{32}{5}\ln2+\frac{83}{90}\right)$.
The anomalous magnetic moment at one-loop in QED under a weak magnetic field is therefore given as 
\begin{equation}
\frac12(g-2)_{\rm QED}^{1-{\rm loop}}=\frac{\alpha}{\pi}\left[\frac12\mp a_2\left(\frac{eB}{m^2}\right)-a_3\left(\frac{eB}{m^2}\right)^2+\cdots\right]\,,
\label{amm_qed}
\end{equation}
where the sign for the odd powers in $B$ depends on the direction of magnetic moment along the magnetic field ($+$ for the parallel and $-$ for the anti-parallel).
Since the energy difference due to the magnetic moment depends on the direction of the magnetic field, the spin-dependent linear term in $B$ in Eq.~(\ref{amm_qed}), which is coming from the $B^2$ correction in the energy difference, Eq.~(\ref{qed_e}), is not measurable in the Penning trap experiment. 

\begin{figure}[t]
\centering
\begin{minipage}[c]{0.32\linewidth}
\centering
  \includegraphics[scale=0.31]{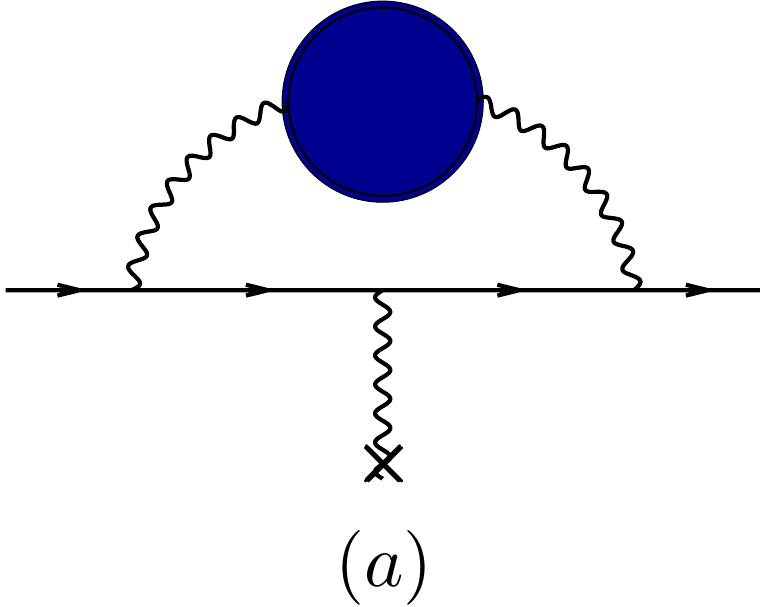}
\end{minipage}
\begin{minipage}[c]{0.31\linewidth}
\centering
  \includegraphics[scale=0.31]{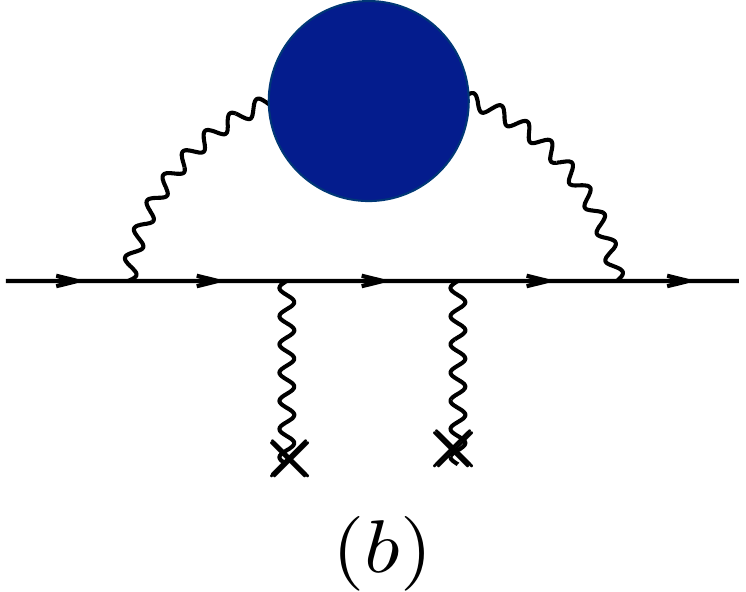}
\end{minipage}
\begin{minipage}[c]{0.31\linewidth}
\centering
  \includegraphics[scale=0.31]{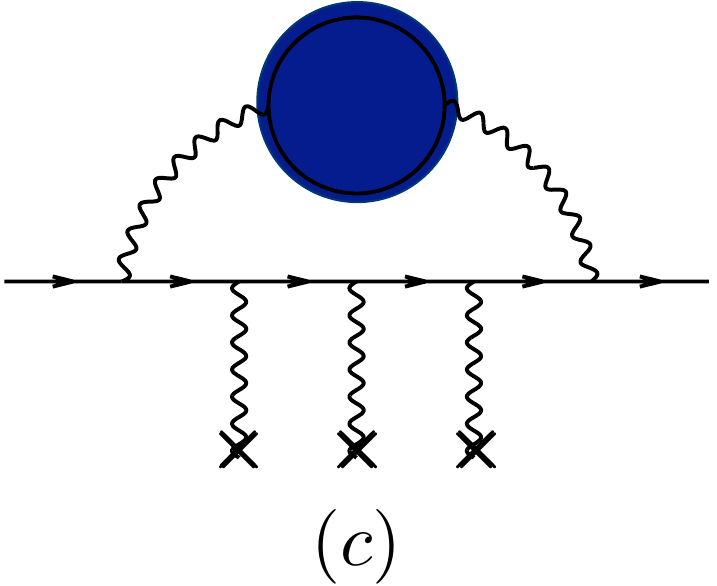}
\end{minipage}
\vskip 0.1in
\begin{minipage}[c]{0.31\linewidth}
\centering
  \includegraphics[scale=0.31]{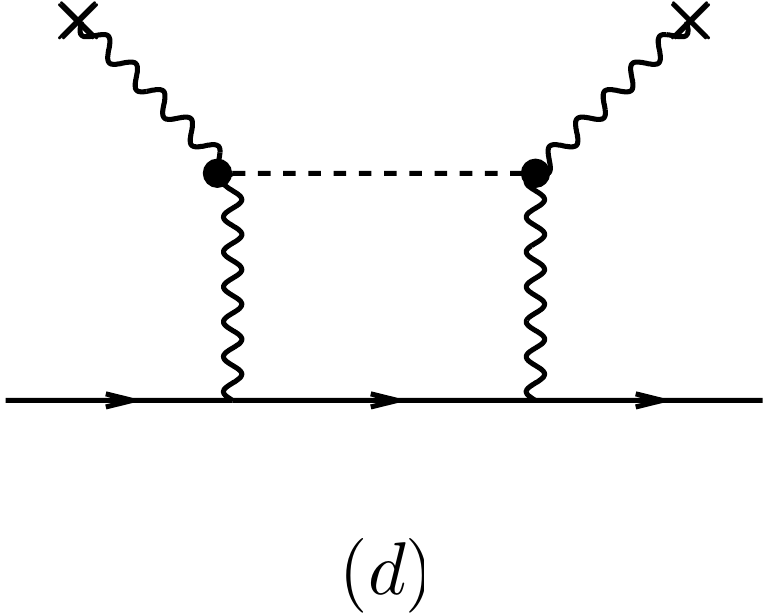}
\end{minipage}
\begin{minipage}[c]{0.31\linewidth}
\centering
  \includegraphics[scale=0.31]{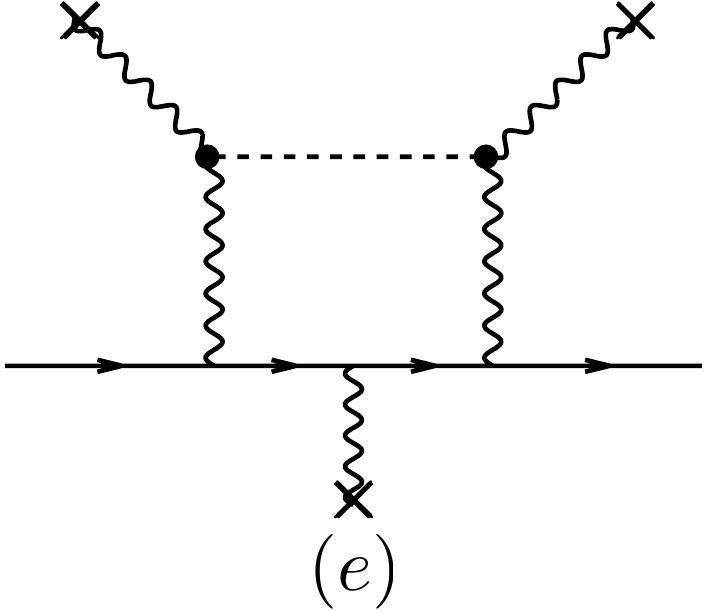}
\end{minipage}

\caption{The hadronic contributions at the order $\alpha^2$ to the ground state energy shift up to the order of $(eB)^3$. The  blob denotes the hadronic vacuum  polarization and the cross denotes the external magnetic field. The solid line denotes the leptons, the wiggly lines photons, and the dashed line the pion.}
 \label{fig1}
\end{figure}

\subsection{Field dependence from QCD corrections}

The QCD corrections to the ground state energy arise through the coupling of photons to QCD correlators.  
The leading QCD corrections involve QCD two-point functions (see Fig.~\ref{fig1}) and are hence of the order $\alpha^2$.
Similarly to the QED  we can therefore expand the leading QCD contribution to the ground state energy as 
\begin{equation}
\Delta{\cal E}^{\rm QCD}(B)=-\left(\frac{\alpha}{\pi}\right)^2\frac{m}{2}\left[c_1\left(\frac{eB}{m^2}\right)+(b_2+c_2)\left(\frac{eB}{m^2}\right)^2+(b_3+c_3)\left(\frac{eB}{m^2}\right)^3+\cdots\right]\,.
\label{qcd_e}
\end{equation}
The QCD corrections to the anomalous moments at the leading order in $\alpha$ is then 
given as 
\begin{equation}
\frac12(g-2)_{\rm QCD}^{\rm LO}=\left(\frac{\alpha}{\pi}\right)^2\left[c_1\pm (b_2+c_2)\left(\frac{eB}{m^2}\right)+(b_3+c_3)\left(\frac{eB}{m^2}\right)^2\cdots\right]\,,
\label{amm_qcd}
\end{equation}
where similarly to Eq.~\eqref{amm_qed} above the $+$ sign is for the anti-parallel magnetic moment to $B$ and the $-$ sign is for the parallel. 
The coefficient $c_1$ gives the field-independent QCD correction at the leading order.
In the large $N_c$ approximation, where the vector mesons dominate, $c_1\sim m^2/m_{\rho}^2$, about $0.019$ for muons. The recent white paper by Muon g-2 Theory Initiative
quotes $c_1(\alpha/\pi)^2=693.1(4.0) \times 10^{-10}$ or $c_1\simeq 0.013$ for muons~\cite{Aoyama:2020ynm}. The coefficients $b_2$ and $b_3$ are coming from the diagrams in Fig.'s~1(b) and 1(c), respectively, while the coefficients $c_2$ and $c_3$ from Fig.'s~1(d) and 1(e), respectively. 

For the leading QCD contribution in $\alpha$, if we calculate up to $(eB)^3$ order, we need to dress the hadronic vacuum polarization contribution with up to three couplings to the external fields, leading to diagrams shown in Fig.~\ref{fig1}.
Notice that due to Furry's theorem, QCD correlators attached to an odd number of photons are forbidden. Apart from the ``standard'' QCD contribution of Fig.~\ref{fig1}(a), we therefore find the diagrams in Fig's.~\ref{fig1}(b) and 1(c) where all external field lines couple directly to the fermion, and contributions where two of the external field lines couple to the QCD part of the diagram. Of these latter diagrams we consider the 
so-called light-by-light diagrams due to hadrons, Fig.'s~1(d) and 1(e), dominated by the exchange of neutral pions with anomalous coupling to photons. The anomalous coupling is given by
\begin{equation}
\int{\rm d}^4xe^{iq_2\cdot x}\left<0\right|{\rm T}j_{\rho}^{\rm em}(x)j_{\sigma}^{\rm em}(0)\left|\pi^0(q_2-q_1)\right>
=iF_{\pi^0\gamma^*\gamma^*}(q_1^2,q_2^2)\,\epsilon_{\mu\nu\rho\sigma}\,q_1^{\mu}q_2^{\nu}\,,
\end{equation}
where $j_{\rho}^{\rm em}(x)$ is the  electromagnetic current of quarks.
The anomalous pion form factor $F_{\pi^0\gamma^*\gamma^*}$ quickly decays when the photon invariant mass becomes bigger than the $\rho$ meson mass or $q^2>m_{\rho}^2$. In this paper we take simply $F_{\pi^0\gamma^*\gamma^*}(q_1^2,q_2^2)\simeq F_{\pi^0\gamma^*\gamma^*}(0,0)=\frac{N_c}{12\pi^2f_{\pi}}$, where $N_c$ is the number of colors and $f_{\pi}=92.4~{\rm MeV}$ is the pion decay constant.

For the diagrams in Fig.'s~\ref{fig1}(b) and 1(c), where the external photon lines attach only to the leptons, we need to calculate the QCD contributions to the vacuum polarization
\begin{equation}
i\int{\rm d}^4xe^{iq\cdot x}\left<0\right|Tj_{\rm em}^{\mu}(x)j_{\rm em}^{\nu}(0)\left|0\right>	=\left(q^2\eta^{\mu\nu}-q^{\mu}q^{\nu}\right)\Pi^{\rm had}_{\rm em}(-q^2)\,,
\end{equation}
where $\eta^{\mu\nu}={\rm diag}(1,-1,-1,-1)$. If we use the vector meson dominance, taking the large $N_c$ limit, we can take for $-q^2\ll m_{\rho}^2$ 
\begin{equation}
	\Pi_{\rm em}^{\rm had}(-q^2)\simeq \frac{f_{\rho}^2}{\left(q^2-m_{\rho}^2\right)}\approx-\frac{f_{\rho}^2}{m_{\rho}^2}\,,
\end{equation}
where the rho meson decay constant $f_{\rho}\approx 0.2\,m_{\rho}$. 
We find then $b_2\simeq 0.04a_2$ and $b_3\simeq 0.04 a_3$. 

The leading $B$-dependent light-by-light contribution to the anomalous magnetic moment, shown in Fig.~\ref{fig1}(d),  depends linearly on the magnetic field. It depends therefore also on the direction of the magnetic moment or the spin of the lepton, which we have already taken into account as the $\pm$ signs in Eq.~\eqref{amm_qcd}. We find, taking $N_c=3$, 
\begin{equation}
c_2= 
\frac{1}{72 \pi^2}\cdot\frac{m^4}{f_{\pi}^2m_{\pi}^2}\left(30\ln\left(\frac{m^2}{m_{\pi}^2}\right)+19\right)\left[1+{\cal O}\left(\frac{m^2}{m_\pi^2}\right)\right]\,.
\end{equation}
From the diagram in Fig.~\ref{fig1}(e) we find the $B^2$ contribution to be (see Appendix A)
\begin{equation}
c_3=-\frac{1}{72 \pi^2}\cdot\frac{m^4}{f_{\pi}^2m_{\pi}^2}\left(12\ln\left(\frac{m^2}{m_{\pi}^2}\right)+17\right)\left[1+{\cal O}\left(\frac{m^2}{m_\pi^2}\right)\right]\,.
\end{equation}

Since the current measurements of the magnetic moments rely on the spin-dependent shift in the Hamiltonian, $\Delta{\cal E}(B)$, it is not possible  to directly measure the leading corrections to the magnetic moment in Eq.'s (\ref{amm_qed}) and (\ref{amm_qcd}) from the energy difference due to the spin flip or from the difference in the spin precession and the cyclotron frequencies. Therefore, the next-to-leading terms,  quadratic in the magnetic field, will affect the measured anomalous magnetic moments, which turn out to be too small for the regular magnetic field to be detected at the experiments of the current accuracy available:
\begin{equation}
\delta g_l(B)\simeq -\frac{2\alpha}{\pi}\left[a_3-\frac{\alpha}{\pi}(b_3+c_3)\right]\cdot\left(\frac{eB}{m^2}\right)^2\simeq1.3\times 10^{-20}\left(\frac{B}{10\,{\rm kG}}\right)^2\left(\frac{0.51\,{\rm MeV}}{m}\right)^4\,.
\end{equation}

\section{Refraction of electrons and the measurement of muon magnetic moment}
\label{sec3}
Since muons are unstable and decay into electrons and neutrinos by the weak interaction with the lifetime of a few $\mu\,{\rm s}$, 
the measurement of the muon magnetic moment has to rely on a completely different method 
from the measurement of the electron magnetic moment~\cite{Combley:1980fg}. 
The current muon anomaly utilizes the fact that the spin precession frequency $\omega_s$ differs from the cyclotron frequency $\omega_c$ when moving in a constant magnetic field. The difference is precisely proportional to the anomalous magnetic moment of muons~\cite{Bargmann:1959gz}:
\begin{equation}
\omega_a\equiv \omega_s-\omega_c=a_{\mu}\frac{eB}{m}\,.
\end{equation}

\begin{figure}[t]
\centering
\begin{minipage}[c]{\linewidth}
\centering
  \includegraphics[scale=0.33]{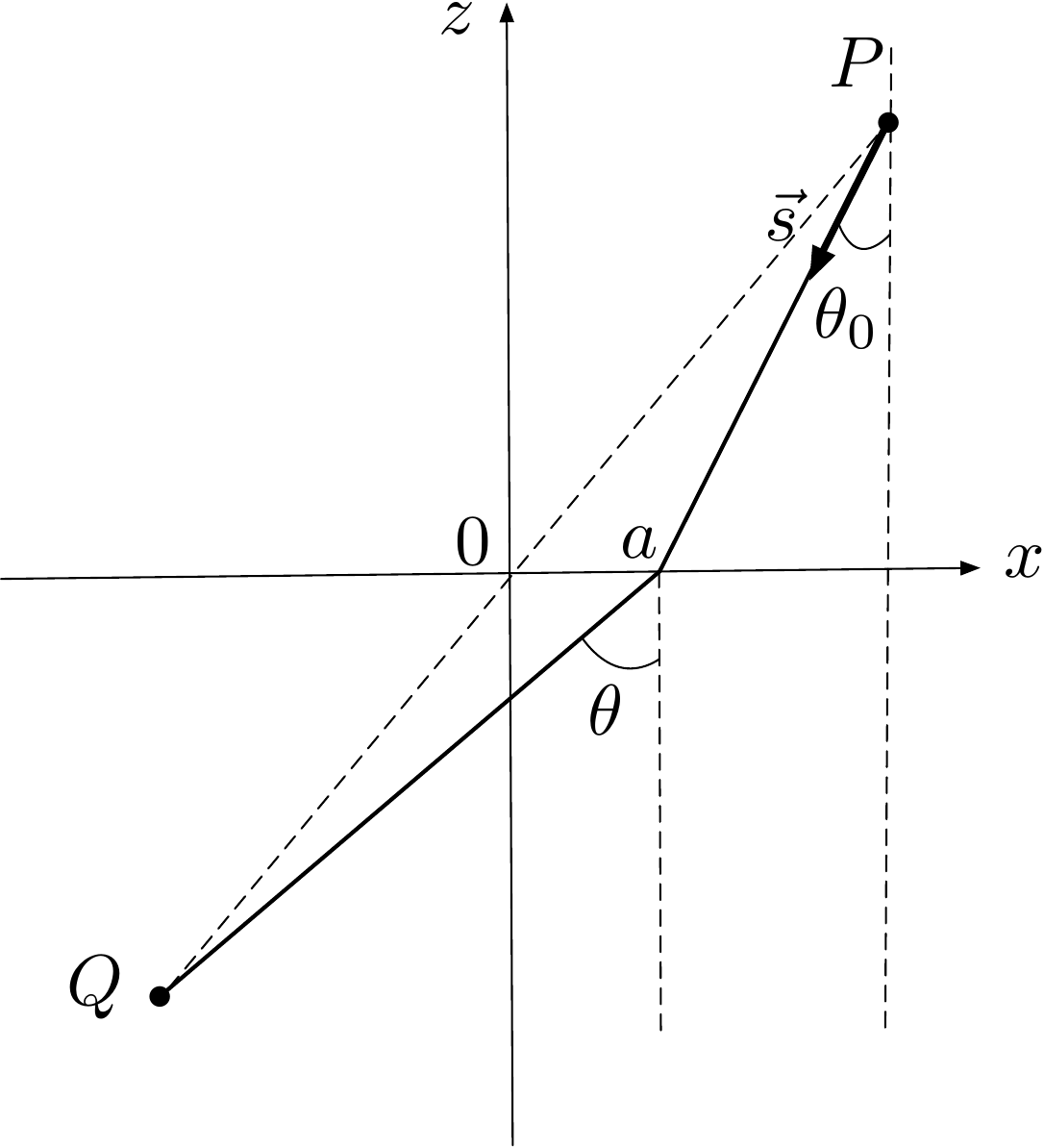}
\end{minipage}

\caption{The top view of the electron path, denoted as a solid line, moving perpendicular to a uniform magnetic field, along the $y$ axis, $\vec B=B{\hat y}$: The electron refracts at $x=a$ on the boundary ($z=0$) of the magnetic field, uniformly distributed in the region $z>0$.  The initial angle $\theta_0=\omega_at+\phi$, where $\phi$ depends on the apparatus geometry. The electron would move along the dotted straight line, if its mass does not change, from $P$ where the muon polarized along $\vec s$ decays, to the electron calorimeter $Q$. }
 \label{fig2}
\end{figure}
If the longitudinally polarized muon decays at a time $t$ after being injected into the magnetic field, the muon spin at the time of decay will be rotated from its momentum direction by an angle $\omega_at$.  As the weak interaction violates the parity, the number of electrons decayed from the muons will be peaked at the muon spin direction. The number of electrons  $N(t)$ detected at the electromagnetic calorimeter will be then given as~\cite{Jegerlehner:2009ry} 
\begin{equation}
N(t)=N_0e^{-t/\tau}\left[1+A\sin\left(\omega_a t+\phi\right)\right]\,,
\label{number}
\end{equation}
where $\tau$ is the boosted lifetime of muon and $\phi$ is the initial misalignment of muon spin, including the geometry of the apparatus. The coefficient $A\simeq (1-2E_e/m_{\mu})/(3-2E_e/m_{\mu})$ is the function of electron energy $E_e$ in units of half muon mass. 
By measuring this oscillatory behavior of the number of electrons detected one can measure the frequency difference and thus the anomalous magnetic moment accurately. However, when the electrons leave the magnetic field to arrive at the calorimeter, they will be refracted since the rest mass of electron will change at the edge of the magnetic field (See Fig.~\ref{fig2}),
\begin{equation}
\delta m\equiv m(B)-m=\Delta {\cal E}(B)\,,
\end{equation}
where $m$ is the electron mass in the absence of the background field and $m(B)$ is the electron rest energy in the background of magnetic field $B$.

To estimate the refracted angle, let us treat the electron semi-classically to follow the least action principle, which is a good approximation since the de Broglie wavelength of electron is much shorter than the typical size of the scale.  In the semi-classical limit the electrons move straight from the decay point $P$ in the orbit of muon to  $x=a$ at the boundary of magnetic field and then exit the magnetic field, moving toward the calorimeter $Q$ from the boundary~\footnote{The electrons will be deflected in the magnetic field by the Lorentz force but this deflection is independent of $\omega_a t$ and therefore can be absorbed into the angle $\phi$.}.
For the constant background field the energy of electron is conserved and therefore the trajectory of the electron is determined in the semi-classical limit by the Maupertuis' principle~\cite{landau}:
\begin{equation}
\delta\int \left(p dl+\vec A\cdot d\vec r\right)=0\,,
\end{equation}
where $dl=\sqrt{d\vec r\cdot d\vec r}$ and $\vec A$ is the vector potential. The momentum satisfies 
\begin{equation}
p^2+m^2=\left({\cal E}-\Delta{\cal E}(B)\right)^2\,.	
\end{equation}
Up to the deflection due to the Lorentz force, the electron trajectory is determined by 
\begin{equation}
\delta\left(\int_P^{x=a}p_1dl+\int_{x=a}^Qp_2dl\right)=0\,.
	\end{equation}
As the momenta in the two regions differ approximately as
\begin{equation}
p_1=p_2\left(1-\frac{{\cal E}\delta m}{p_2^2}\right)\,,	
\end{equation}
we find the refracted angle $\theta$ is related to the incident angle $\theta_0=\omega_at+\phi$ in the leading order in $\delta m/m$ as
\begin{equation}
\tan\theta-\tan\theta_0=\gamma_*\tan\theta_0\cdot \sec^2\theta_0\cdot\frac{\delta m}{m}\,,
\label{angle}
\end{equation}
where $\gamma_*=m{\cal E}/p_2^2\approx m/p_2$.
When $\tan\theta_0$ is small, 
the increment in the refraction angle $\delta\theta=\theta-\theta_0$ is~\footnote{Right after the muon decays the electrons are almost longitudinally polarized but the magnetic moments are constantly measured upon scattering with the external magnetic field. The electrons arrived at the calorimeter are  polarized either up or down along the $B$ field direction. Hence the displacement position at the boundary in Fig.~\ref{fig2} is either $-a$ for up or $+a$ for down with equal probability.}
\begin{equation}
\delta\theta=-\gamma_*\frac{\delta m}{m}\tan\theta_0\sim 10^{-16}\left(\frac{1~{\rm GeV}}{p_2}\right)\left(\frac{B}{10\,{\rm kG}}\right)\tan\theta_0\,.
\end{equation}
For a generic angle $\theta_0$, the sine of the refracted angle is given from Eq.~(\ref{angle}) as
\begin{equation}
\sin\theta=\frac{\sin\theta_0\left(\cos^2{\theta_0}+\gamma_*\delta m/m\right)}{
\sqrt{\cos^6\theta_0+\sin^2\theta_0\left(\cos^2{\theta_0}+\gamma_*\delta m/m\right)^2}}\,.
\end{equation}
The oscillatory part  in the formula in Eq. (\ref{number}) should be therefore given by $\sin\theta$ instead of $\sin\theta_0$, taking account of the refraction by the magnetic field.
\begin{figure}[t]
\centering
\begin{minipage}[c]{0.45\linewidth}
\centering
  \includegraphics[scale=0.25]{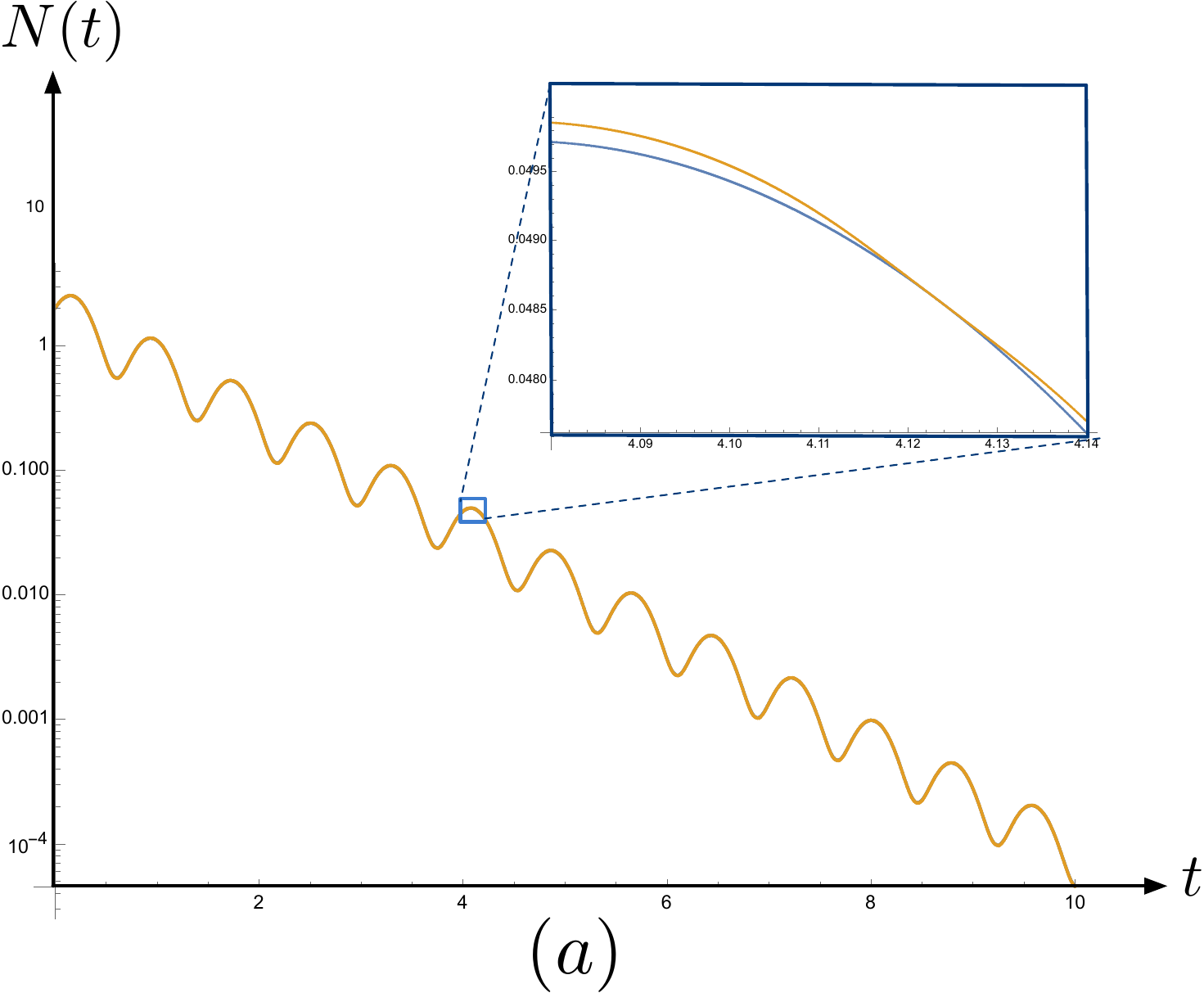}
\end{minipage}
\hskip 0.1in
\begin{minipage}[c]{0.45\linewidth}
\centering
  \includegraphics[scale=0.30]{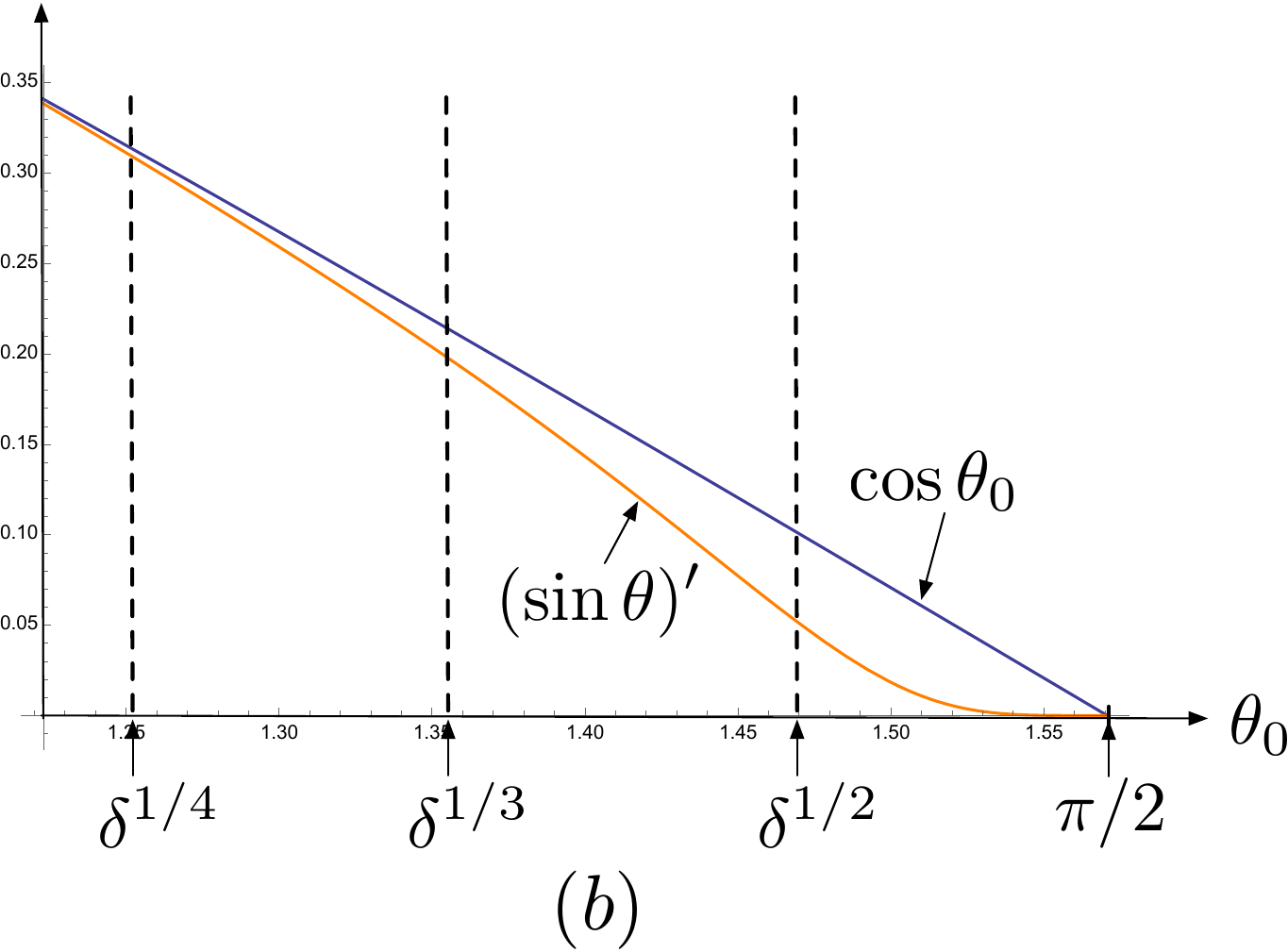}
\end{minipage}
\caption{(a) Distribution of electron counts versus time in arbitrary scales, taking $A=0.5$ and $\gamma_*\delta m/m=0.01$. The distribution near the 6th peak at $t=4.12$ in the blue box is shown enlarged. The orange curve is the refracted distribution with $\sin\theta$ and the blue one is the distribution with $\sin\theta_0$, not considering the refraction effect. (b) The comparison of the slope the oscillatory parts of two distributions near $\pi/2$. The prime denotes the derivative with respect to $\theta_0$ and $\delta^n\equiv\pi/2-(\gamma_*\delta m/m)^n$.}
\label{fig3}
\end{figure}
We find that the amplitude of the oscillatory part of the distribution is amplified roughly by $\gamma_*\delta m/m$ except at  
peaks, valleys, and nodes, $\theta_0=n\pi/2$ $(n=1,2,3,\cdots)$, where two distributions coincide, $\sin\theta=\sin\theta_0$: 
\begin{equation}
{\gamma_*}\frac{\delta m}{m}\gtrsim\left|\sin\theta-\sin\theta_0\right|\ge0\,.
\end{equation}
 The maximum of the difference, $\sin\theta-\sin\theta_0$, is $\gamma_*\delta m/m$ that occurs 
at $\Delta\theta_0\simeq (3\gamma_*\delta m/m)^{1/4}$ away from the peaks and valleys. We see that the slope of $\sin\theta$ is rather flat near the peaks and valleys for $\Delta\theta_0\simeq(3\gamma_*\delta m/m)^{1/4}$. The slopes differ drastically   at $\pi/2-(\gamma_*\delta m/m)^{1/2}$ or near the peaks or valleys (See Fig.~\ref{fig3}(b)). 
Since in the current Fermilab Muon $g-2$ experiment, about $10^9$ positrons are detected in each run subset, about $10^5$ positrons will then arrive for $\Delta\theta_0\sim 10^{-4}$, statistically enough to differentiate two distributions with $\sin\theta$ or $\sin\theta_0$.

Due to the refraction, the distribution of electrons changes notably near the peaks and valleys, $\theta_0=n\pi/2$ $(n=1,3,5\cdots)$ (See Fig.~\ref{fig3}(a))~\footnote{The peak (valley) occurs at the angle $\theta_0=\pi/2 \,(3\pi/2)$, namely when the electrons leave almost tangentially to the boundary of the magnetic field.}. 
We note that the refraction, however, does not change the frequency of the electron distribution but changes only its shape. 
Especially the slope of the distribution of the detected positrons (or electrons) will be quite flat near the peaks and valleys, when the refraction effect is included. With the statistics of the current muon $g-2$ experiment,  one could therefore measure the refraction effect due to the magnetic field by fitting the distribution function, especially its slope near the peaks and valleys, while improving the uncertainties in the estimate of the muon anomaly, $a_{\mu}$.

\section{Conclusion}
\label{con}
We have examined the effect of  magnetic fields on the magnetic moment of leptons. 
As the magnetic moments of electrons and muons are being measured with an extreme accuracy, the effect of magnetic fields may not be negligible. For the pure QED contribution one can read off the effect
from the known results for the ground state energy under the constant magnetic field~\cite{Newton:1954zz,Tsai:1973jex}
while for the QCD corrections we calculate the leading hadronic contributions up to quadratic order in the external magnetic field, which consist of the leading hadronic vacuum polarization as well as the light-by-light scattering. 

One naively expects the effect of the magnetic field would be $\frac{\alpha}{\pi}|eB|/m^2$ and comparable to the current experimental uncertainties for electrons. However, the effect turns out to come from the quadratic order in $(eB)$ and thus too small to be detected.  We also estimate the refraction of the electron at the boundary of the magnetic field, which is found to be linear in $B$ but quadratic in $\alpha$ with the kinematical suppression, $m{\cal E}/p^2$, due to the large electron momentum $p$. The refraction changes the shape of the electron distribution at the electron counter at the level of $10^{-8}$, which is quite small but could be measurable at the current muon $g-2$ experiment. With the modified shape one could not only measure the refraction effect but also improve the experimental uncertainties in the measurement of muon $g-2$.

Finally in the Appendices we discuss the general field-dependent form factors and we show the Ward-Takahashi identity in the presence of the magnetic field, needed to check the consistency of our calculations.

\acknowledgments
This work was supported by the National Research Foundation of Korea (NRF) grant funded by the Korea government (MSIT) (2021R1A4A5031460) and also by Basic Science Research Program through the National Research Foundation of Korea (NRF) funded by the Ministry of Education (NRF-2017R1D1A1B06033701) (DKH). The work of T.~D. and M.~J. has been supported in part by an appointment to the JRG Program at the APCTP through the Science and Technology Promotion Fund and Lottery Fund of the Korean Government. T.~D. and M.~J. have also been supported by the Korean Local Governments -- Gyeongsangbuk-do Province and Pohang City -- and by the National Research Foundation of Korea (NRF) funded by the Korean government (MSIT) (grant number 2021R1A2C1010834).

\appendix
\label{app}
\section{Field-dependent form factors}

Recall that the standard expression for the amputated photon-fermion interaction vertex (in the absence of external fields) has the form
\begin{equation}\label{eq:B0formf}
    \Gamma^\mu(p',p) = \gamma^\mu F_1(q^2) + \frac{i\sigma^{\mu\nu} q_\nu}{2m} F_2(q^2)
\end{equation}
so that there are only two terms which are consistent with Lorentz covariance, hermiticity, discrete symmetries, and the Ward identity 
\begin{equation}
    \bar u(p')q_\mu \Gamma^\mu(p',p)u(p)  = 0 \,.
\end{equation}
We denoted here $q=p'-p$.

\subsection{Form factors at $\mathcal{O}(F^2)$}

Including interactions with an external gauge field, described in terms of a constant field strength tensor $F_{\mu\nu}$, however gives rise to several additional possible structures. We consider these structures at $\mathcal{O}(F^2)$. Notice that the Ward identity is no longer that simple in this case because the interactions with the external field give rise to nontrivial self energy (see Appendix~\ref{app:WT}). For brevity, we will limit ourselves to the ``parity even'' terms, i.e., terms free of $\gamma_5$. We have checked that the ``parity odd'' terms do not contribute to $g-2$ for the diagrams of Fig.~\ref{fig1}.

Let us first write down the covariant factors at $\mathcal{O}(F^2)$ where the nontrivially Lorentz transforming piece does not contain the external fields. These include the  $\mathcal{O}(F^2)$ corrections to the terms of~\eqref{eq:B0formf} as well as the additional term
\begin{equation}
      (p'-p)^\mu \left[{F_\beta}^\alpha F^{\beta\rho}(p_\alpha p_\rho - p'_\alpha p'_\rho)\right]\,,
\end{equation}
which is no longer forbidden by the Ward identity.  If we assume only magnetic field in the rest frame, i.e., that
\begin{equation} \label{eq:Brestf}
  (p'+p)_\mu F^{\mu\nu} = 0    
\end{equation}
this term vanishes.

There are also several Lorentz covariant, nontrivially transforming terms which can we written down by a adding a single occurrence of $F^{\mu\nu}$. At $\mathcal{O}(F^2)$ they are multiplied by the only nonzero scalar $\mathcal{O}(F)$ term $p_\rho F^{\rho\beta}p'_\beta$. The possible terms are
\begin{alignat}{3}
    &1 &\times&F^{\mu\nu}(p-p')_\nu&\times&\left[p_\rho F^{\rho\beta}p'_\beta\right]\,,\\
    &\gamma^\alpha&\times&  F_{\alpha\nu}(p+p')^\nu (p'-p)^\mu&\times&\left[p_\rho F^{\rho\beta}p'_\beta\right]\,,\\
            &\gamma^\alpha &\times & F_{\alpha\nu}(p-p')^\nu (p+p^{\prime})^{\mu}&\times&\left[p_\rho F^{\rho\beta}p'_\beta\right]\,,\\
    &i\sigma^{\alpha\nu} &\times & F_{\alpha\nu}(p+p')^\mu &\times&\left[p_\rho F^{\rho\beta}p'_\beta\right]\,,\\
    &i\sigma^{\alpha\mu} &\times& F_{\alpha\nu}(p+p')^\nu&\times&\left[p_\rho F^{\rho\beta}p'_\beta\right] \,.
\end{alignat}
All these terms again vanish if we assume only magnetic field in the rest frame because then $p_\rho F^{\rho\beta}p'_\beta = 0$.

Finally one can also write down terms which involve two $F$ tensors in a nontrivial manner. 
The terms which vanish under the condition~\eqref{eq:Brestf} are
\begin{alignat}{2}
    &1 &\times&F^{\mu\nu}F_{\nu\alpha}(p+p')^{\alpha}\,, \label{eq:FFpterm}\\
    &\gamma_\alpha &\times&F^{\mu\beta}F^{\alpha\nu}(p+p')_\beta(p+p')_\nu\,, \\
    &\gamma_\alpha &\times&F^{\alpha\beta}F_{\beta\nu}(p+p')^\mu(p+p')^\nu\,, \\
    &i\sigma_{\alpha\beta} &\times & F^{\alpha\nu} F^{\beta\rho}(p-p')_\nu(p+p')_\rho(p+p')^\mu \,.
\end{alignat}
There are also 
other terms that do not vanish under this condition, but are of higher order in $q=p'-p$, not contributing to $g-2$:
\begin{alignat}{2}
    &\gamma_\alpha &\times&F^{\mu\beta}F^{\alpha\nu}(p-p')_\beta(p-p')_\nu \label{eq:FFpp1}\,,\\
    &\gamma_\alpha &\times&F^{\alpha\beta}F_{\beta\nu}(p-p')^\mu(p-p')^\nu \label{eq:FFpp2}\,.
\end{alignat}
This leaves us fives terms that may potentially contribute:  
\begin{alignat}{2}
    &\gamma^\alpha &\times&F_{\alpha\beta}F^{\beta\mu}\,, \label{eq:FFterm}\\
    &i\sigma^{\alpha\mu} &\times&F_{\alpha\beta}F^{\beta\nu}(p-p')_\nu \,,\label{eq:sigmaterm}\\
        &i\sigma^{\alpha\beta} &\times & F_{\alpha\beta}F^{\mu\nu}(p-p')_\nu\,, \\
    &i\sigma_{\alpha\beta} &\times & F^{\alpha\mu}F^{\beta\nu}(p-p')_\nu\,, \\
    &i\sigma^{\alpha\beta} &\times & {F_{\alpha}}^\nu F_{\beta\nu}(p-p')^\mu \,.\label{eq:sigmamod} 
\end{alignat}
To summarize, there are (in the general case) 19 ``parity even'' form factors appearing at  $\mathcal{O}(F^2)$. Seven of them, i.e., the corrections to $F_1$ and $F_2$ as well as the form factors of~\eqref{eq:FFterm}--\eqref{eq:sigmamod}, may contribute to $g-2$.

\subsection{Correction to $g-2$}

In the rest of the Appendix we will assume the limit $q \to 0$ and only keep the terms arising from the diagram in Fig.~\ref{fig1}~(e). As it turns out, this leaves us only with the correction to the $F_2$ and the terms in~\eqref{eq:FFterm} and in~\eqref{eq:sigmaterm}.\footnote{Naively there is also an IR divergent correction to $F_1$, but we expect that the physical effect from this term is zero following the same arguments as for the leading order $F_1$ contribution.}
We may write the result at $\mathcal{O}(F^2)$ as
\begin{equation}
    \delta \Gamma^\mu \approx 
    \frac{i\hat c_1\sigma^{\mu\nu}}{2m}q_\nu F^2 +\frac{i\hat c_2\sigma^{\alpha\nu}}{2m} F_{\alpha\beta}F^{\beta\mu}q_\nu+  \frac{i\hat c_3\sigma^{\alpha\mu}}{2m} F_{\alpha\beta}F^{\beta\nu}q_\nu\,,
\end{equation}
where the coefficients $\hat c_i$ are independent of the momenta but may depend on masses and we used Gordon identity to 
rewrite~\eqref{eq:FFterm} in terms of $\sigma^{\mu\nu}$.

As we have isolated the small terms in $p-p'$, we can take a simple rest frame where (apart from the explicit factors of $p-p'$) the three-momenta of the fermions are zero. Plugging in $F^{ij} = - \epsilon^{ijk}B^k$ 
we obtain
\begin{align} \label{eq:vertexq0}
    -i\bar u(p')  \delta \Gamma^k u(p) &\approx -\left(4\hat c_1-2\hat c_2+2\hat c_3\right) B^2 \epsilon^{kln} q^l S^n & \nonumber\\
    & \ \ - 2\hat c_2 B^k \epsilon^{lmn}B^lq^mS^n  +2\hat c_3 \epsilon^{klm}B^lS^m \, q^nB^n\,,
\end{align}
where the spin is
\begin{equation}
 S^k = -\frac{1}{8m}   \bar u(p') \epsilon^{ijk}\sigma^{ij} u(p) \ .  
\end{equation}

To  derive the final result, we should couple to the external (constant) magnetic field. The magnetic field is generated by a vector potential, which we may take to be
\begin{equation} \label{eq:vectorpot}
    A^k(\vec x) = - \frac{1}{2} \epsilon^{klm} x^l B^m \ .
\end{equation}
Taking the Fourier transform leads to 
\begin{equation}
    q^l\tilde A^k(\vec q) =  \frac{i}{2} \epsilon^{klm}B^m (2\pi)^3 \delta^3(\vec q) \,,
\end{equation}
which is antisymmetric under the exchange $l \leftrightarrow k$. 
Inserting this in~\eqref{eq:vertexq0} gives
\begin{align}
    \bar u(p')  \delta \Gamma^k u(p)A^k(\vec q) &\approx \left(4\hat c_1-2\hat c_2+2\hat c_3\right) B^2\,\vec B \cdot \vec S\ (2\pi)^3 \delta^3(\vec q) \,.
\end{align}

The values of the coefficients $\hat c_i$ can be extracted from the computation of the diagram in Fig.~\ref{fig1}~(e), which we carried out by using package X for Mathematica~\cite{Patel:2015tea}: 
\begin{align}
    \hat c_1 &= \frac{7 c_a^2 e^2}{24 \pi ^2 m_\pi^2}+O\left(m^2\right) & \\
    \hat c_2 &=\frac{c_a^2 e^2 \left(12 \log \left(\frac{m^2}{m_\pi^2}\right)+23\right)}{72 \pi ^2m_\pi^2}+O\left(m^2\right) & \\
    \hat c_3 &=-\frac{c_a^2 e^2}{2 \pi ^2 m_\pi^2}+O\left(m^2\right) & \\
    4 \hat c_1-2\hat c_2+2\hat c_3&= -\frac{c_a^2 e^2 \left(12 \log \left(\frac{m^2}{m_\pi^2}\right)+17\right)}{36 \left(\pi ^2 m_\pi^2\right)}+O\left(m^2\right)\,,&
\end{align}
where the anomaly coefficient, $c_a=N_c e^2/(12 \pi^2 f_{\pi}) = \alpha/(\pi f_{\pi})$ with $N_c=3$. 
In the limit of small fermion mass, the contribution from the $\hat c_2$ term dominates due to the logarithm.

\section{The Ward-Takahashi identity under magnetic field}\label{app:WT}
The Ward-Takahashi (WT) identity is a diagrammatic manifestation of gauge symmetry of the theory. It is useful to consider it as a consistency check of our calculation of the diagram in Fig.~\ref{fig1}(e).

In simple terms, the WT identity states the following: A diagram involving an external photon can be expressed as a  sum of simpler diagrams which lack of the external photon but otherwise identical to original one. To derive the former from the latter, one should just sum over all the simpler diagrams with all the possible insertions of the photon. For our case, the WT identity is 
\begin{equation}
q_\mu \, \delta \Gamma^\mu(p+q,p)  =e\left(\Sigma(p)-\Sigma(p+q)\right),
\end{equation}
where $q^\mu$ is the momentum of the external photon, the $\delta$'s refer to contributions involving a pion exchange with two anomaly vertices and the self-energy contribution $\Sigma$ (See Fig.~\ref{fig1}(d)) is 
\begin{eqnarray}
\Sigma(p)=e^2\int \frac{d^4k}{(2\pi)^4}\gamma_{\alpha}\frac{(\slashed p-\slashed k+m)}{(p-k)^2-m^2}\gamma_{\beta} \Pi^{\alpha\beta}(k).
\end{eqnarray}
The result for r.h.s. of the identity is 
\begin{eqnarray}
\left(\Sigma(p)-\Sigma(p+q)\right)&=&d_1 F_{\alpha\beta}F^{\alpha\beta}+d_2 \gamma_\alpha q_\beta {F_\gamma}^\alpha F^{\gamma\beta}+d_3 p_\alpha p_\beta {F_{\gamma}}^{\alpha}F^{\gamma\beta}\nonumber\\
&&+d_4 (2 q_\alpha q_\beta {F_\gamma}^\alpha F^{\gamma\beta}+ p_\alpha q_\beta {F_\gamma}^\alpha F^{\gamma\beta}),
\end{eqnarray}
where the coefficients $d_i$ are, assuming that the fermion momenta are on-shell,
\begin{eqnarray}
d_1&=&
\slashed q\frac{\left(8 m^6/{m^2_\pi}+2 m^2 {m^2_\pi}-4 m^4\right) \Lambda\left(m^2,m,m_\pi\right)}{12 m^6}\nonumber\\
&&+\slashed q\frac{2 m^2 {m^2_\pi}+\left(-4 m^2 {m^2_\pi }+6 m^4+{m^4_\pi}\right) 2\log \left(m/{m_\pi }\right)-5 m^4}{12 m^6},
\end{eqnarray}
\begin{eqnarray}
d_2&=&
\frac{2 \left(2 m^4 {m_\pi^2}-m^2 {m_\pi^4}+8 m^6\right) \Lambda\left(m^2,m,{m_\pi }\right)}{6 m^6 {m_\pi^2}}\nonumber\\
&&+\frac{{m_\pi^2} \left(-2 m^2 {m_\pi^2}+\left(4 m^2 {m_\pi^2}+6 m^4-{m_\pi^4}\right) 2\log \left(m/{m_\pi}\right)+5 m^4\right)}{6 m^6 {m_\pi^2}},
\end{eqnarray}
\begin{eqnarray}
d_3&=&
-\frac{\left(-20 m^4 {m_\pi^2}+6 m^2 {m_\pi^4}+8 m^6\right) \Lambda\left(m^2,m,{m_\pi}\right)}{6 m^8 {m_\pi^2}}\nonumber\\
&&-\frac{{m_\pi^2} \left(6 m^2 {m_\pi^2}+\left(-16 m^2 {m_\pi^2}+18 m^4+3 {m_\pi^4}\right) 2\log \left(m/{m_\pi}\right)-23 m^4\right)}{6 m^8 {m_\pi^2}},
\end{eqnarray}
\begin{eqnarray}
d_4&=&
\frac{\Lambda \left(m^2,m,{m_\pi}\right) \left(20 m^4 {m_\pi^2} \slashed p+\slashed q)-6 m^2 {m_\pi^4} (\slashed p+\slashed q)-8 m^5 {m_\pi^2}-8 m^6 (\slashed p+\slashed q)+8 m^7\right)}{12 m^8 {m_\pi^2}}\nonumber\\
&&-\frac{{m_\pi^2} \left(2 \log (m/{m_\pi}) \left(-16 m^2 {m_\pi^2} (\slashed p+\slashed q)+4 m^3 {m_\pi^2}+18 m^4 (\slashed p+\slashed q)-12 m^5+3 {m_\pi^4} (\slashed p+\slashed  q)\right)\right)}{12 m^8 {m_\pi^2}}\nonumber\\
&&-\frac{{m_\pi^2} \left(6 m^2 {m_\pi^2} (\slashed p+\slashed q)-23 m^4 (\slashed p+{\slashed q})+8 m^5\right)}{12 m^8 {m_\pi^2}},
\end{eqnarray}
where $\Lambda(s,m_0,m_1)$ is the part of the Passarino-Veltman $B_{0}$ function containing the $s$ plane branch cut.

We conclude this appendix by remarking that above result equals to $q^\mu$-contracted and amputated  diagram (See Fig.~\ref{fig1}(e)) involving pion exchange with two anomalous vertices. The above check is performed on-shell, which is sufficient for the calculation of $g-2$. The off-shell check is also possible, but we do not pursue it here.  

\end{document}